\documentclass[option,numerical,linenumbers,superscriptaddress,nofootinbib,showpacs,showkeys,longbibliography]{webofc}
\usepackage[varg]{txfonts} 
\usepackage{graphicx,color}
\usepackage{amsmath,amssymb,amsfonts}
\usepackage[colorlinks=true,linkcolor=blue,plainpages=false,hypertex]{hyperref}

\def \be {\begin{equation}}
\def \ee {\end{equation}}
\def \ee  {\end{equation}}
\def \bea {\begin{eqnarray}}
\def \eea {\end{eqnarray}}

\def \roots{\mbox{$\sqrt{s_{_{NN}}}$}\xspace}

\newcommand{\pT} {\ensuremath{p_{\mathrm{T}}}}
\begin{document}
\title{Beam energy and system dependence of rapidity-even dipolar flow}
\author{\firstname{Niseem Magdy} \lastname{(For the STAR Collaboration)}\inst{1}\thanks{\email{niseemm@gmail.com}}}
\institute{Department of Chemistry, Stony Brook University, Stony Brook, NY, 11794-3400, USA}
\abstract{
New measurements of rapidity-even dipolar flow, v$^{even}_{1}$, are presented for several transverse momenta, $p_T$, and centrality intervals in Au+Au collisions at $\sqrt{s_{NN}}~=~200,~39$ and $19.6$ GeV, U+U collisions at $\sqrt{s_{NN}}~=~193$ GeV, and Cu+Au, Cu+Cu, d+Au and p+Au collisions at $\sqrt{s_{NN}}~=~200$~GeV. 
The v$^{even}_{1}$ shows characteristic dependencies on $p_{T}$, centrality, collision system and $\sqrt{s_{_{NN}}}$, consistent with the expectation from a hydrodynamic-like expansion to the dipolar fluctuation in the initial state. These measurements could serve as constraints to distinguish between different initial-state models, and aid a more reliable extraction of the specific viscosity $\eta/s$.
}
\maketitle
\section{Introduction}
\label{intro}

Heavy-ion collisions (HIC) at the Relativistic Heavy Ion Collider (RHIC) and the Large Hadron Collider (LHC) 
are aimed at studying the properties of the strongly interacting quark-gluon plasma (QGP) created in such collisions. Recent studies have emphasized the use of anisotropic flow measurements to study the transport properties of the QGP ~\cite{Teaney:2003kp,Lacey:2006pn,Schenke:2011tv,Song:2011qa,Niemi:2012ry,Qin:2010pf,Magdy:2017kji}.  A crucial question in these studies was the role of initial-state fluctuations and their influence on the uncertainties associated with the extraction of $\eta/s$ for the QGP produced in HIC ~\cite{Alver:2010gr,Lacey:2013eia}. This work emphasizes new measurements for rapidity-even dipolar flow, v$^{even}_{1}$, which could aid a distinction between different initial-state models and facilitate the extraction of $\eta/s$ with better constraints.

Anisotropic flow is characterized by the Fourier coefficients,  v$_{n}$, obtained from a Fourier expansion of the azimuthal angle ($\phi$) distribution of the emitted particles ~\cite{Poskanzer:1998yz}:
\begin{eqnarray}
\label{eq:1}
\frac{dN}{d\phi}\propto1+2\sum_{n=1}\mathrm{v_{n}}\cos (n(\phi-\Psi_{n})),
\end{eqnarray}
where $\Psi_n$ represents the $n^{th}$-order event plane, the coefficients v$_{1}$, v$_{2}$ and v$_{3}$ are called directed,  elliptic and  triangular flow, respectively. The flow coefficients  v$_{n}$ are related to the two-particle Fourier coefficients v$_{n,n}$ as:
\begin{eqnarray}
\label{eq:3}
\mathrm{v_{n,n}}(\pT^{a},\pT^{b})  = \mathrm{v_n}(\pT^{a})\mathrm{v_n}(\pT^{b})+ \delta_{NF},
\end{eqnarray}
where $\pT^{a}$ and $\pT^{b}$ are the transvers momentum of particles (a) and (b), respectively, and $\delta_{NF}$ is a so-called  non-flow (NF) term, which  includes possible contributions from resonance decays, Bose-Einstein correlations, jets, and global momentum conservation (GMC) ~\cite{Lacey:2005qq,Borghini:2000cm,Luzum:2010fb,Retinskaya:2012ky,ATLAS:2012at}. 
The directed flow, v$_{1}$, can be separated into an odd function of pseudorapidity ($\eta$) ~\cite{Danielewicz:2002pu} which develops along the direction of the impact parameter, and a rapidity-even component ~\cite{Teaney:2010vd,Luzum:2010fb} which results from the effects of initial-state fluctuations acting in concert with a hydrodynamic-like expansion; v$_1(\eta) =$ v$^{even}_1(\eta) +$ v$^{odd}_1(\eta)$, where $\Psi^{odd}_1$ and $\Psi^{even}_1$ are uncorrelated. The magnitude of v$^{even}_1$ is related to the fluctuations-driven  dipole asymmetry $\varepsilon_1$ and $\eta/s$ ~\cite{Teaney:2010vd,Gardim:2011qn,Retinskaya:2012ky}.

\section{Measurements}
\label{Measurements}
The correlation function technique was used to generate the two-particle $\Delta\phi$ correlations:
\begin{eqnarray}\label{corr_func}
 C_{r}(\Delta\phi, \Delta\eta) = \frac{(dN/d\Delta\phi)_{same}}{(dN/d\Delta\phi)_{mixed}},
\end{eqnarray} 
where  $(dN/d\Delta\phi)_{same}$ represent the normalized azimuthal distribution of  particle pairs  from the same 
event and $(dN/d\Delta\phi)_{mixed}$ represents the normalized azimuthal distribution for particle pairs
in which each member  is selected  from a different  event but with a similar classification for the vertex,  centrality, etc. The pseudorapidity requirement  $|\Delta\eta| > 0.7$ was also imposed on track pairs to minimize possible non-flow contributions associated with the short-range correlations from resonance decays, Bose-Einstein correlations and jets.
%
%
\begin{figure*}[tb]
\centering{
\includegraphics[width=0.7\linewidth,angle=0]{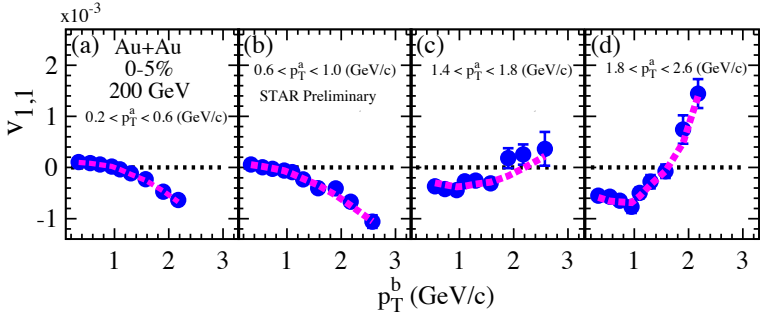}
\vskip -0.3cm
\caption{v$_{1,1}$ vs. $p_{T}^{b}$ for several selections of $p_{T}^{a}$  for 
0-5\% central Au+Au collisions at $\sqrt{s_{_{NN}}} = 200$~GeV. The dashed curve shows the 
result of the simultaneous fit with Eq.~\ref{corrv1}.
 \label{fig:1}
 }
}
\vskip -0.6cm
\end{figure*}

The two-particle Fourier coefficients v$_{n,n}$ are obtained from the correlation function as:
\begin{eqnarray}\label{vn}
\mathrm{v_{n,n}} &=& \frac{\sum_{\Delta\phi} C_{r}(\Delta\phi, \Delta\eta)\cos(n \Delta\phi)}{\sum_{\Delta\phi}~C_{r}(\Delta\phi, \Delta\eta)},
\end{eqnarray}
and  then used to extract v$^{even}_{1}$ via a simultaneous fit of v$_{1,1}$ as a function of $p_{T}^{\text {b}}$, for several 
selections of  $p_{T}^{a}$ with Eq.~\ref{eq:3}:
\begin{eqnarray}\label{corrv1}
\mathrm{v_{1,1}}(\pT^{a},\pT^{b})  &=& \mathrm{v^{even}_{1}}(\pT^{a})\mathrm{v^{even}_{1}}(\pT^{b}) - C\pT^{a}\pT^{b}.
\end{eqnarray}
Here, $C \propto 1/(\langle Mult \rangle \langle p_{T}^{2}\rangle)$ takes into account the non-flow correlations induced by a global momentum conservation~\cite{Retinskaya:2012ky,ATLAS:2012at} and $\langle Mult \rangle$ is the mean multiplicity.

For  a given centrality selection, the left hand side of  Eq.~\ref{corrv1} represents the $N \times N$ matrix which we fit with the right hand side using $N + 1$ parameters; N values of v$^{even}_{1}(\pT)$ and one additional parameter $C$, accounting for momentum conservation~\cite{Jia:2012gu}.  
Fig.~\ref{fig:1} shows a representative result for this fitting procedure for $0-5\%$ central Au+Au collisions at $\roots = 200$~GeV. The dashed curve (obtained with Eq.~\ref{corrv1}) in each panel illustrates the effectiveness of the simultaneous fits, as well as the constraining power of the data. That is, v$_{1,1}(\pT^{b})$ evolves from negative to positive values as the selection range for  $\pT^a$ is increased.

\section{Results}
\label{Results}
\begin{figure*}[tb]
\centering{
\includegraphics[width=0.95\linewidth,angle=0]{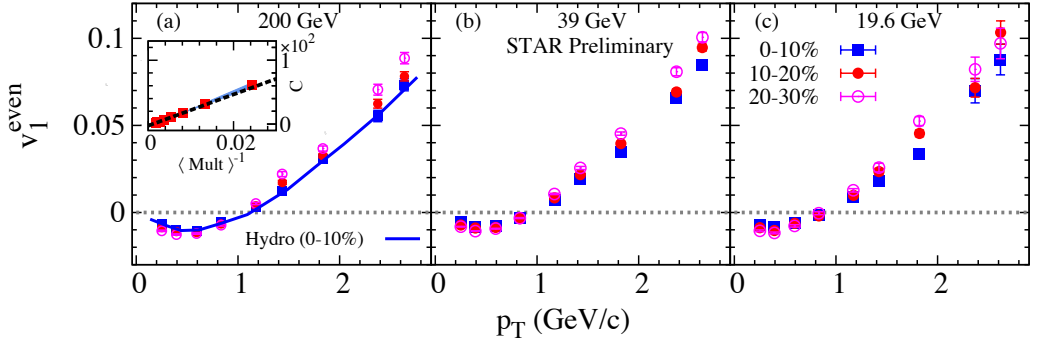}
\caption{Extracted values of v$^{even}_{1}$ vs. $p_T$ for different centrality selections (0-10\%, 10-20\% and 20-30\%) Au+Au collisions for several values of  $\sqrt{s_{_{NN}}}$ as indicated; the v$^{even}_{1}$ values are obtained via fits with Eq.~(\ref{corrv1}). The solid line in panel (a) shows the result from a hydrodynamic calculations with $\eta/s~=~ 0.16$ \cite{Retinskaya:2012ky}.  The inset in panel (a) shows a representative set of the associated values of $C$ vs. $\langle Mult\rangle^{-1}$.
 \label{fig:2}
 }
}
\vskip -0.4cm
\end{figure*}

\begin{figure*}[tb]
\centering{
\includegraphics[width=0.95\linewidth,angle=0]{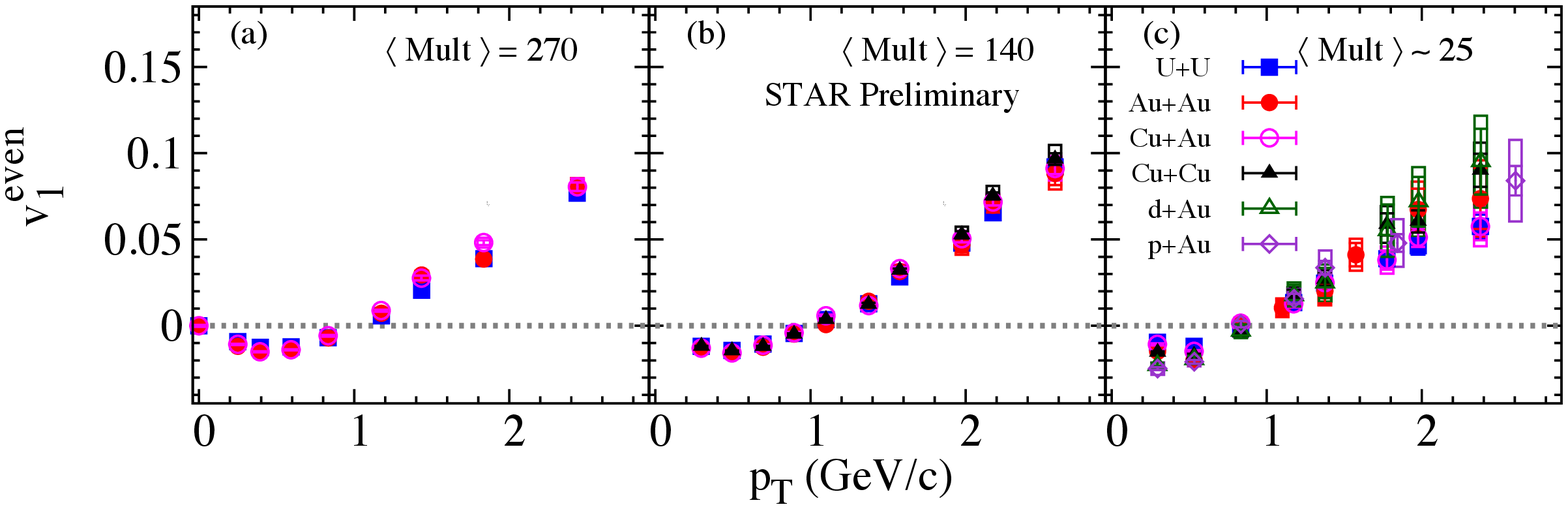}
\vskip -0.3cm
\caption{Extracted values of v$^{even}_{1}$ vs. $p_T$ for different $\langle Mult \rangle$ selections for different collisions system at  $\sqrt{s_{_{NN}}} \sim 200$ GeV as indicated; the v$^{even}_{1}$ values are obtained via fits with Eq.~(\ref{corrv1}). 
 \label{fig:3}
 }
}
\vskip -0.6cm
\end{figure*}

Representative v$^{even}_{1}$ results for Au+Au collisions at $\sqrt{s_{NN}}~=~ 200,~39,$ and $19.6$ GeV 
and for different collision systems U+U at $\sqrt{s_{NN}}~=~193$ GeV, and Cu+Au, Cu+Cu, d+Au 
and p+Au at $\sqrt{s_{NN}}~=~200$ GeV are summarized in Figs.~\ref{fig:2} and~\ref{fig:3}.
%
%
%
%
The values of v$^{even}_{1}(\pT)$ extacted for different centrality selections (0-10\%, 10-20\% and 20-30\%) 
are shown  in Fig.~\ref{fig:2}; the solid line in panel (a) shows the a hydrodynamic calculations with $\eta/s~=~ 0.16$\cite{Retinskaya:2012ky}, which in good agreement with our measurements, the inset shows the corresponding results for the associated momentum conservation coefficient, $C$, extracted for several centralities at $\roots = 200$~GeV. The v$^{even}_{1}(\pT)$  values indicate the characteristic pattern of a change from negative  v$^{even}_{1}(\pT)$ at low $\pT$ to positive v$^{even}_{1}(\pT)$ for $\pT > 1$ ~GeV/c, with a crossing point that shifts with $\roots$. They also indicate that v$^{even}_{1}$ increase as the centrality become more peripheral, as might be expected from the centrality dependence of $\varepsilon_1$.

The extracted values of v$^{even}_{1}(\pT)$,  for different collision systems are compared 
in Fig.~\ref{fig:3} for different values of $\langle Mult \rangle$. Figs.~\ref{fig:3}(a), \ref{fig:3}(b) 
and \ref{fig:3}(c) indicate similar v$^{even}_{1}(\pT)$ magnitudes for the systems specified at 
each $\langle Mult \rangle$, as well as the characteristic pattern of a change from negative v$^{even}_{1}(\pT)$ at low $\pT$ to positive v$^{even}_{1}(\pT)$ for $\pT > 1$~GeV. 
This pattern confirms the predicted trends for rapidity-even dipolar flow ~\cite{Teaney:2010vd,Luzum:2010fb,Retinskaya:2012ky} and further indicates that for the selected  values of $\langle Mult \rangle$,  v$^{even}_{1}(\pT)$ does not show a strong dependence on the collision system.  This apparent system independence of  v$^{even}_{1}(\pT)$ for the indicated $\langle Mult \rangle$ values suggests that the fluctuations-driven initial-state eccentricity $\varepsilon_1$, is similar for the six collision systems. It also suggests that the viscous effects that are related to $\eta/s$ are comparable for the matter created in each of these collision systems.

\section{Conclusion}
\label{Conclusion}

In summary, we have used the two-particle correlation method to carry out new differential measurements of 
rapidity-even dipolar flow, v$^{even}_{1}$, in Au+Au collisions at different beam energies, and in U+U, Cu+Au, Cu+Cu, d+Au and p+Au collisions at $\roots \simeq 200$~GeV. The measurements confirm the characteristic patterns of  an evolution from negative v$^{even}_{1}(\pT)$  for $\pT >1$~GeV/c to positive v$^{even}_{1}(\pT)$ for $\pT > 1$~GeV/c, expected when initial-state geometric fluctuations act in concert with the hydrodynamic-like expansion to generate rapidity-even dipolar flow. This measurements provide additional constraints which are important to discern between different initial-state models, and to aid precision extraction of the temperature dependence of the specific shear viscosity.

\section*{Acknowledgments}
This research is supported by the US Department of Energy under contract DE-FG02-87ER40331.A008.

%
\bibliography{ref_BES_v1}

\end{document}